\documentstyle[10pt]{article}
\bf
 \newcommand{\hs}[1]{\hspace*{ #1 mm}}
 \newcommand{\vs}[1]{\vspace*{ #1 mm}}

 \newcommand{\setempty}{\mathrm{\O}}

 \newcommand{\prob}{{\mathrm{Prob}}}

 \newcommand{\etalc}{\textrm{et al.}}

 \newenvironment{proof}{\par \noindent
            {\bf Proof. \hs{2}}}{\hfill$\Box$ \vspace*{3mm}}

 \newcommand{\qubit}[1]{| #1 \rangle}

 \newcommand{\qsc}{\mathrm{QSC}}
 \newcommand{\dsc}{\mathrm{DSC}}
 \newcommand{\psc}{\mathrm{PSC}}
 \newcommand{\twoqfa}{\mathrm{2QFA}}

 \newcommand{\polytime}{poly\mbox{-}time}

\newcommand{\ignore}[1]{}

\newcommand{\cent}{{|}\!\!\mathrm{c}}

\begin{document}
\title{Research report:\\  {\bf
State complexity of operations on two-way quantum finite
automata}}
\author{Daowen Qiu\thanks{
Email-address: issqdw@mail.sysu.edu.cn }\\
{\footnotesize Department of Computer Science, Zhongshan
University, Guangzhou, 510275, P.R. China}}
\date{  }
\maketitle
 {\bf Abstract}
\par
\vskip 1mm  This paper deals with the size complexity of minimal
{\it two-way quantum finite automata} (2qfa's) necessary for
operations to perform on all inputs of each fixed length. Such a
complexity measure, known as state complexity of operations, is
useful in measuring how much information is necessary to convert
languages. We focus on intersection, union, reversals, and
catenation operations and show some upper bounds of state
complexity of operations on 2qfa's. Also, we present a number of
non-regular languages and prove that these languages can be
accepted by 2qfa's with one-sided error probabilities within
linear time. Notably, these examples show that our bounds obtained
for these operations are not tight, and therefore worth improving.
We give an instance to show that the upper bound of the state
number for the simulation of one-way deterministic finite automata
by two-way reversible finite automata is not tight in general.

\vskip 2mm

\smallskip

\section*{1. Introduction}

\subsection*{1.1.  Background of this topic and relevant results}

Finite (state) automata represent one of the most simple form of
computation to date. The minimal number of inner states used in a
finite automaton has been a focal point of a recent study of
finite automata (for example, see [5,9,13,15,16,26,30] and the
references therein). Such a number serves as a complexity measure
and is generally referred to as {\em state complexity}. The
importance of the study of state complexity arises from recent
applications of finite automata to numerous practical fields,
which include natural language and speech processing, software
engineering, and image generation and encoding, etc [7,25]. In the
literature, there are by and large three distinct streams of
studies concerning state complexity of finite automata.
\begin{itemize}
\item State complexity of language recognition on a fixed automata model.

\item State complexity of transformation of one automata model to another.

\item State complexity of operations on a fixed automata model.
\end{itemize}

Indeed, these complexity measures have a long history of their
own. In the 1950s, Rabin and Scott [24] showed that an $n$-state
one-way nondeterministic finite automaton (1nfa) is transformed to
its equivalent one-way deterministic finite automaton (1dfa) of at
most $2^n$ inner states. This is generally referred to as the
state complexity of transformation. The notion of state complexity
of operations, on the contrary, dates back to 1981 when Leiss [18]
showed that the reversal of a langauge recognized by an $n$-state
1dfa requires $2^n$ inner states.

In this paper, we focus on the state complexity of operations on
{\it two-way quantum finite automata} (2qfa's), which is defined
as the minimal number of inner states necessary for a finite
automaton to ``witness'' a given operation on all inputs of each
fixed length. An automaton that achieves such minimality is called
a {\em minimal automaton}.

Since the state complexity dealt with in this report is concerning
quantum computing models, we recall some background regarding
quantum computing. Exactly, quantum mechanical computation---a new
breed of nature-inspired computation---has recently drawn wide
attention as an alternative computation paradigm [12,20]. The
notion of quantum finite automata (qfa's) was introduced by the
pioneering work of Moore and Crutchfield [19] and Kondacs and
Watrous [17]. The computational model of Kondacs and Watrous is a
natural quantization of probabilistic finite automata, which have
been studied for more than four decades since its introduction by
Rabin [24]. Their model consists of an input tape, a read-only
head, and a finite control unit holding an inner state and evolves
by a rule of quantum mechanics. As two-way quantum finite
automata, the read-only head moves along the input tape in two
directions (also being allowed to stay still). A configuration of
such a machine is in general in a so-called superposition. The
models of Kodacs and Watrous and of Moore and Cruchfield differ in
one point: the number of measurements performed during a
computation. Specifically, Kodacs and Watrous' models perform
measurement each step of computing, whereas Moore and
Crutchfield's ones only measure at the end of a computation.

Note that, similar to probabilistic finite automata, 2qfa's and
1qfa's are quite different in power. We are particularly
interested in the power of two-way quantum finite automata
(2qfa's). Kondacs and Watrous [17] showed that, by exploiting
quantum interference, 2qfa's can recognize even non-regular
languages, in particular, $Upal=\{0^n1^n\mid n\in {\bf N}\}$ in
worst-case linear time. This shows a sharp contrast with the fact
that there is a regular language that cannot be recognized by any
one-way quantum finite automaton (1qfa) with any constant error
bound because of its reversibility constraint. Notably, Amano and
Iwama [2] showed that the empty problem for certain restricted
2qfa's (which they call 1.5qfa's) is undecidable.

It is worth mentioning that Ambainis, ect. [1,3], Nayak [21],
Brodsky and Pippenger [6], and the others have dealt with some
operation properties and state size on quantum finite automata.
For example, Ambainis etc. [3] proved that the union of the
languages accepted by one-way quantum finite automata with bounded
error is not closed.

\subsection*{1.2. Our goals and obtained results}

A recent interest in practical fields makes state complexity a
practically important and theoretically interesting entity. It is
natural to discuss state complexity based on quantum finite
automata. Is such complexity measure quite different from that
classical one? In this paper, we focus our study on the state
complexity of operations on fixed automata models, i.e., 2qfa's,
that were first proposed and studied by Kodacs and Watrous [17].
We attempt to demonstrate that this complexity measure indeed
proves vital in quantum complexity theory.

For notational convenience, we introduce the notation
$QSC_{\epsilon}[L](n)$ to denote the smallest number of inner
states necessary to solve language $L$ on a 2qfa model $M$ with
error probability bounded above by $\epsilon$ when inputs are
exactly length $n$. To be more precise, for any $x\in L\cap
\Sigma^{n}$, the probability of accepting $x$ is at least
$1-\epsilon$, and the probability of rejecting $x\in
\overline{L}\cap \Sigma^{n}$ is also at least $1-\epsilon$, where
$\Sigma^n$ denotes the set of strings with length $n$, and
$\overline{L}$ is the complement of $L$. For simplicity, we call
{\it $M$ accepting $L\cap\Sigma^{n}$ with error probability
bounded by $\epsilon$}.

In this paper, we mainly prove the following results.

{\bf Theorem 1.} For any languages $L_{1}$ and $L_{2}$ over
$\Sigma_{1}$ and $\Sigma_{2}$ respectively, and any $n\in {\it
N}$, let $M_{1}$ and $M_{2}$ be the minimum 2qfa for
$L_{1}\cap\Sigma_{1}^{n}$ and $L_{2}\cap\Sigma_{2}^{n}$ with error
probabilities bounded by $\epsilon_{1}$ and $\epsilon_{2}$,
respectively. If $M_{2}$ is non-recurrent, then
\begin{equation}
QSC_{\epsilon}[L_{1}\cap L_{2}](n)\leq
QSC_{\epsilon_{1}}[L_{1}](n)+QSC_{\epsilon_{2}}[L_{2}](n)\times
(n+2)\times |Q_{acc,1}|-|Q_{acc,1}|
\end{equation}
\begin{equation}
QSC_{\epsilon}[L_{1}\cup L_{2}](n)\leq
QSC_{\epsilon_{1}}[L_{1}](n)+QSC_{\epsilon_{2}}[L_{2}](n)\times
(n+2)\times |Q_{rej,1}|-|Q_{rej,1}|
\end{equation}
where
$\epsilon=\epsilon_{1}+\epsilon_{2}-\epsilon_{1}\epsilon_{2}$,
$|Q_{acc,1}|$ denotes the number of the rejecting states of
$M_{1}$.

{\bf Theorem 2.} For any language $L$ over $\Sigma$, let $M$ be
the minimum 2qfa for $L$ with error probability bounded by
$\epsilon$. If $M$ is non-recurrent, then

\begin{equation}
QSC_{\epsilon}[L](n)-1\leq QSC_{\epsilon}[L^R](n)\leq
QSC_{\epsilon}[L](n)+1
\end{equation}
for any $n\in {\bf N}$, where $L^R$ denotes the reversal of $L$,
i.e., $L^R=\{x^R|x\in L\}$ (for
$x=\sigma_{1}\sigma_{2}\ldots\sigma_{n}$, then
$x^R=\sigma_{n}\sigma_{n-1}\ldots\sigma_{1}$).

{\bf Theorem 3.} Let $L_{i}$ be a language over alphabet
$\Sigma_{i}$ with $\epsilon\not\in L_{i}$ for $i=1,2$. If
$\Sigma_{1}\cap\Sigma_{2}=\emptyset$, and the error probabilities
of the minimum 2qfs's $M_{1}$ and $M_{2}$ accepting $L_{1}$ and
$L_{2}$ are respectively $\epsilon_{1}$ and $\epsilon_{2}$, then
the catenation $L_{1}L_{2}$ is accepted by a 2qfa $M$ with error
probability
$\epsilon=\epsilon_{1}+\epsilon_{2}-\epsilon_{1}\epsilon_{2}$.

{\bf Proposition 1.} For alphabet $\Sigma=\{a,b_{1},b_{2}\}$, let
$L=\{a^{n}b_{1}^{n}a^mb_{2}^m: n,m\geq 1\}$. Then $L$ is accepted
by 2qfa with one-sided error in linear time.

{\bf Proposition 2.} For alphabet $\Sigma=\{a,b_{1},b_{2}\}$, let
$L_{1}=\{a^{+}b_{1}^{+}a^mb_{2}^m: m\geq 1\}$ and
$L_{2}=\{a^{n}b_{1}^{n}a^+b_{2}^+: n\geq 1\}$. Then there exist
2qfa $M_{1}$ and $M_{2}$ accepting $L_{1}$ and $L_{2}$,
respectively, with one-sided error in linear time.

{\bf Proposition 3.} There exists regular language $L$ satisfying
\begin{equation}
QSC_{0}[L](2n)\leq \frac{\sqrt{DSC_{0}[L](2n)}}{6}-4
\end{equation}
for any $n\in {\bf N}$, where $DSC_{0}[L](n)$ denotes the smallest
number of inner states necessary to accept language $L$ on a
one-way deterministic finite automaton when inputs are exactly
length $n$.

From the above theorems and propositions it follows a number of
corollaries which will be stated in Sections 3 and 4. Also, in
order to prove the above results, we need verify some lemmas, that
will be detailed in the sequel.

\section*{2. Basic notions and notation}

We review some related notions and notation that will be used in
this paper. In addition, some will be explained when they appear
first.

\subsection*{2.1. General definitions}

Let ${\bf N}$ be the set of all natural numbers (that is,
nonnegative integers), ${\bf Z}$ be that of all integers, ${\bf
R}$ be that of all real numbers, and ${\bf C}$ be that of all
complex numbers. Let ${\bf N}^{+}={\bf N} - \{0\}$. For any
complex number $\alpha$, $\alpha^*$ denotes its conjugate. For any
two numbers $m,n\in{\bf N}$ with $m<n$, the notation $[m,n]_{\bf Z
}$ denotes the set $\{m,m+1,m+2,\ldots,n\}$. For any finite set
$Q$, $|Q|$ denotes the cardinality of $Q$.

We use the notation $\Sigma$ in general for a nonempty input
alphabet (not necessarily be limited to $\{0,1\}$). A {\em string
over $\Sigma$} is a finite sequence of symbols in $\Sigma$ and the
{\em length} of a string is the number of occurrences of symbols
in the string. In particular, the string of length $0$ is called
the {\em empty string} and often denoted $\lambda$. For each
number $n\in {\bf N}$, $\Sigma^n$ denotes the set of all strings
over $\Sigma$ that have length exactly $n$. Write $\Sigma^*$ for
$\bigcup_{n\in{\bf N}}\Sigma^n$. A {\em partial problem} over
alphabet $\Sigma$ is a pair $(A,B)$ such that $A,B\subseteq
\Sigma^*$ and $A\cap B=\empty$. When $A\cup B=\Sigma^*$, $B$
becomes the complement of $A$ (denoted $\Sigma^*-A$) and thus we
identify $(A,B)$ with $A$, which is simply called a {\em
language}.

We assume the reader's familiarity with classical finite automata.
We use the notation $\mathrm{REG}$ to denote the collection of all
regular languages.

Let $A$ be any $m\times n$ complex matrix. The notation $A^T$
denotes the transposed matrix of $A$. Moreover, $A^{\dagger}$
denotes the transposed conjugate of $A$. For any vector $x$,
$\|x\|$ denotes the $\ell_2$-norm. Let $\|A\|$ be the operator
norm defined as $\|A\|=\max\{\|Ax\|: \|x\|\neq0\}$. Let
$\|A\|_{2}$ be the Frobenius norm
$\left(\sum_{i,j}|a_{i,j}|^2\right)^{1/2}$. Let $\|A\|_{tr}$ be
the trace norm $\min\{|Tr(AX^{\dagger})|: \|X\|\leq 1\}$.

\subsection*{2.2. Quantum finite automata}

We briefly give the formal definition of a quantum finite
automaton (qfa). Formally, a qfa is described as a sextuple
$M=(Q,\Sigma,q_0,\delta,Q_{acc},Q_{rej})$, where $Q$ is a finite
set of inner states with $Q_{acc}\cup Q_{rej}\subseteq Q$,
$\Sigma$ is a finite alphabet, $q_0$ is the initial inner state,
and $\delta$ is a transition function mapping from $Q\times\Sigma$
to $Q\times\{-1,0,+1\}$. The transition function $\delta$ is also
expressed by a series of {\em transitions} define by the complexy
number (called an amplitude) $\delta(p,\sigma,q,d)$ for $p,q\in
Q$, $\sigma\in\Sigma$, and $d\in\{-1,0,+1\}$. This means that,
assuming that a machine is in inner state $p$ scanning a symbol
$\sigma$, the machine at next step changes its inner state to $q$
moving its head in direction $d$. The set $Q$ is partitioned into
three sets: $Q_{acc}$, $Q_{rej}$, and $Q_{non}$. Inner states in
$Q_{acc}$ (in $Q_{rej}$, resp.) are called {\em accepting states}
({\em rejecting states}, resp.). A {\em halting state} refers to
both an accepting state and a rejecting state. The rest $Q_{non}$
is known as a set of {\em non-halting states}. A {\em
configuration} is a description of a single moment of $M$'s
computation, including an inner state and a head position; we
regard $Q\times[0,|x|+1]_{\bf Z}$ as the configuration space on
input $x$. In general, $M$'s computation is a series of
superpositions of configurations, each of which evolves by an
application of $\delta$ to its predecessor (if not the initial
configuration). More generally, we can view an application of
$\delta$ as an application of a linear operator over a
configuration space. More precisely, the  operator $U_{\delta}^n$
with respect to inputs of length $n$ is defined as the linear
operator acting on the configuration space ${\cal CONF}_{n}$,
which is the Hilbert space spanned by $\{\qubit{p,i}\mid q\in
Q,i\in[0,n+1]_{\bf Z}\}$, in the following way: for each $(p,i)\in
Q\times[0,n+1]_{\bf Z}$, let $U^x_{\delta}\qubit{p,i} =
\sum_{(q,d)\in
Q\times\{0,\pm1\}}\delta(p,x_i,q,d)\qubit{q,i+d\,(\mathrm{mod}\,n+2)}$,
where $x_0=\cent$, $x_{n+1}=\$$, and $x_i$ is the $i$th symbol of
$x$ for each $i\in[1,n]_{\bf Z}$. Throughout this paper, we assume
that $U_{\delta}^n$ is always unitary for any $n\in {\bf N}$.

We say that a 2qfa $M$ {\em recognizes $(A,B)$ with error
probability $\epsilon$} if (i) for every $x\in A$, $M$ accepts $c$
with probability $1-\epsilon$ and (ii) for every $x\in B$, $M$
rejects $x$ with probability $1-\epsilon$. When $(A,B)$ is
identified with the language $A$, we simply say that $M$
recognizes $A$. For notational simplicity, we write
$\prob_{M}[M(x)=1]$ to denote the probability of $M$ accepting
$x$. Similarly, $\prob_{M}[M(x)=0]$ for the probability of $M$
rejecting $x$. We define the class $\twoqfa$ as the collection of
all languages that can be recognized by 2qfa's with error
probability at most certain constant $\epsilon\in[0,1/2)$.
Similarly, $\twoqfa(\polytime)$ is defined by 2qfa's which run in
expected polynomial time.

\section*{3. State complexity by quantum finite automata}

We formally introduce the notation of {\em state complexity} of
langauge recognition on quantum finite automata.

\subsection*{3.1. Definition of state complexity}

Roughly speaking, the state complexity of a language is the number
of inner states of the minimal qfa that recognizes the language
with designated error probability. More formally, we say that a
language $L$ over alphabet $\Sigma$ {\em has state complexity}
$s(n)$ with error probability $\epsilon$ and amplitude set $K$ if
there exists a 2qfa
$M_n=(Q_n,\Sigma,q_0,\delta_n,Q_{acc,n},Q_{rej,n})$ such that (i)
$|Q_n|\leq s(n)$ and (ii) for every $x\in L\cap \Sigma^n$,
$\prob_{M_n}[M_n(x)=1]= 1-\epsilon$, and (iii) for every $x\in
\overline{L}\cap\Sigma^n$,  $\prob_{M_n}[M_n(x)=0]= 1-\epsilon$.
When $n$ is concerned, we say that $L$ has state complexity $s(n)$
at $n$ with error probability $\epsilon$. In the literature, qfa's
are sometimes specified by their partial transitions because it is
easy to expand such partial transitions to standard transitions
(see [27,28,29], for example). To discuss the number of inner
states, it is therefore important to note that here we consider
only ``complete'' qfa's, which are equipped with transition
functions defined completely on the domain $Q\times\Sigma$.

For notational convenience, we write $\qsc_{K,\epsilon}[L](n)$ for
the minimal number $m$ such that $L$ has state complexity $m$ at
$n$ with error probability at most $\epsilon$ and with amplitude
set $K$. In particular, when $K={\bf C}$, we omit the subscript
$K$ from $\qsc_{K,\epsilon}[L](n)$ for readability. For
comparison, we also introduce the notations for classical state
complexity measures. Write $\dsc[L](n)$ for the minimal number of
inner states of any 2dfa recognizing $L$. A {\em minimal 2qfa}
refers to a 2qfa that witnesses $\qsc_{K,\epsilon}[L](n)$.
Moreover, we write $\psc_{\epsilon}[L](n)$ for the minimal number
of inner states of any 2pfa recognizing $L$ with error probability
at most $\epsilon$. Since any 2dfa is also a 2pfa, we clearly
obtain the following relationships: Let $L$ be any language and
let $\epsilon\in[0,1/2)$. $\dsc[L](n)\geq \psc_{\epsilon}[L](n)$
for all $n\in {\bf N}$. If the running time of automata are
concerned, we write $\qsc_{K,\epsilon}^{poly}[L](n)$
($\qsc_{K,\epsilon}^{lin}[L](n)$, resp.), for instance, to
emphasize that we use only 2qfa's which run in expected polynomial
time (expected linear time, resp.).

In the subsequent subsection, we review basic properties of state
complexity defined by 2qfa's.

\subsection*{3.2. Fundamental properties}

We show several fundamental properties of state complexity
measures.

{\bf Lemma 1.} Let $L$ be any language over alphabet $\Sigma$, let
$\epsilon$ be any constant in $[0,1/2)$, and let $K$ be any
amplitude set.
\begin{enumerate}\vs{-2}
\item $1\leq \qsc_{\epsilon}[L](n)\leq \sum_{i=0}^{n}|\Sigma|^{i}$ for any $n\in {\bf N}$.
\item $\qsc_{0}^{lin}[L](n)\leq 2\dsc[L](n)+6$ for all $n\in {\bf N}$.
\end{enumerate}

\begin{proof}
1) The lower bound comes from the fact that every 2qfa requires
the initial inner state and at least one accepting or rejecting
inner state. Actually, we can set $Q_{acc}=\{q_0\}$ and
$Q_{rej}=\setempty$ in the case of $L=\Sigma^*$. In this case,
$\qsc_{0}[\Sigma^*](n)=1$ for all $n\in {\bf N}$. The upper bound
is shown as follows. After the machine reads the left end-marker
symbol, the initial state evolves also into initial state. Then
the machine needs at most $|\Sigma|$ states after scanning the
first input symbol (since there are only $|\Sigma|$ symbols). For
the second step, there are at most $|\Sigma|$ states and each
evolves at most $|\Sigma|$ states, which implies that the second
step needs at most $|\Sigma|^{2}$ states. Proceeding with this
method, we therefore get the above result.

2) By the proof of Proposition 4 in the reference [17] by Kondacs
and Watrous, we directly obtain this result.
\end{proof}

A natural problem is whether or not the bound in Lemma 2 is tight.
Indeed, from the following Lemma 2 we know that for some language
$L$ we have Proposition 3 stated in Section 1. Therefore it says
that it is not tight.

{\bf Lemma 2.} Let $\Sigma=\{a,b\}$, and let $L=\{w\in\Sigma^{*}|
\#_{a}(w)=\#_{b}(w)=n\}$, where $n\in {\bf N}$, and $\#_{a}(w)$
and $\#_{b}(w)$ denote respectively the number of $a$ and $b$ in
the string $w$. Then there exists 2qfa (exactly 2rfa, i.e.,
two-way reversible finite automaton) $M$ with $6n+4$ states to
accept $L$.

\begin{proof}
We construct a 2rfa $M=(Q,\Sigma,\delta,q_{0},Q_{acc},Q_{rej})$ as
follows.

$Q=\{q_{0},q_{1},\ldots,q_{2n+1},p_{1},p_{2},\ldots,p_{n+1},p_a,
q_{r}^{(1)},q_r^{(2)},\ldots,q_r^{(2n)},p_r^{(1)},p_r^{(2)},\ldots,p_r^{(n)}\}$,

$Q_{acc}=\{p_{a}\}$,
$Q_{rej}=\{q_{r}^{(1)},q_r^{(2)},\ldots,q_r^{(2n)},p_r^{(1)},p_r^{(2)},\ldots,p_r^{(n)}\}$,

$\delta(p,\sigma,q,d)=\left\{\begin{array}{ll}\langle
q|V_{\sigma}|p\rangle, & D(q)=d,\\
0,& {\rm otherwise},\end{array} \right.$ for any $p,q\in Q$ and
any $\sigma\in \Sigma\cup \{\cent,\$\}$, where unitary operators
$V_{\sigma}: \ell_{2}(Q)\rightarrow \ell_{2}(Q)$ and mapping
$D:Q\rightarrow \{0,\pm 1\}$ are defined as:

$V_{\cent}|q_{0}\rangle=|q_{1}\rangle$,
$V_{\cent}|p_{n+1}\rangle=|p_{a}\rangle$,
$V_{\cent}|p_{i}\rangle=|p_{\sigma}^{(i)}\rangle$,
$i=1,2,\ldots,n$,

$V_{\$}|q_{2n+1}\rangle=|p_{1}\rangle$;\\
for $\sigma\in \Sigma$,

$V_{\sigma}|q_{i}\rangle=|q_{i+1}\rangle$ for $i=1,2,\ldots,2n$,
$V_{\sigma}|q_{2n+1}\rangle=|q_{r}^{(1)}\rangle$,

$V_{\sigma}|p_{i}\rangle=|p_{i+1}\rangle$ for $i=1,2,\ldots,n$,

$V_{a}|p_{n+1}\rangle=|p_{r}^{(1)}\rangle$. \\ Then $V_{\sigma}$
is readily extended unitarily to $\ell(Q)$.

$D(q_{i})=1$ for $i=0,1,\ldots,2n+1$, and $D(q)=0$ for any $q\in
Q_{acc}\cup Q_{rej}$.

One can readily check that the 2rfa constructed above can accept
$L=\{w\in\Sigma^{*}| \#_{a}(w)=\#_{b}(w)=n\}$ with $6n+4$ states.
\end{proof}

However, in terms of the Myhill- Nerode Theorem [14], any
deterministic finite automaton accepting $L$ needs at least $n^2$
states. Therefore, for language $L$ above we have that Proposition
3 holds.

{\bf Proposition 3.} There exists regular language $L$ satisfying
\begin{equation}
QSC_{0}[L](2n)\leq \frac{\sqrt{DSC_{0}[L](2n)}}{6}-4
\end{equation}
for any $n\in {\bf N}$.

The following lemma is immediate from the definitions of
$\twoqfa$, $\twoqfa(\polytime)$, and state complexity.

{\bf Lemma 3.} Let $L\subseteq\Sigma^*$.
\begin{enumerate}\vs{-2}
\item $L\in \twoqfa$ iff there exist constants $\epsilon'\in[0,1/2)$ and $c\in{\bf N}$ such that $\qsc_{\epsilon'}[L](n)\leq c$ for all $n\in {\bf N}$.
\vs{-2}
\item $L\in \twoqfa(\polytime)$ iff there exist constants
$\epsilon'\in[0,1/2)$ and $c\in{\bf N}$ such that
$\qsc_{\epsilon'}^{poly}[L](n)\leq c$ for all $n\in{\bf N}$.
\vs{-2}
\end{enumerate}

\section*{4. State complexity of basic operations}

The main theme of this paper, as stated in Section 1, is to
determine the upper bounds of state complexity of basic operations
on 2qfa's. First, we consider the operation called
``complementation'' on 2qfa's. This is the easiest case. {}From
the definition of 2qfa's, we immediately obtain the exact bound as
follows.

{\bf Lemma 4.} For any language $L$ and any constant
$\epsilon\in[0,1/2)$, $\qsc_{\epsilon}[\Sigma^*-L](n) =
\qsc_{\epsilon}[L](n)$ for all $n\in{\bf N}$.

\begin{proof}
This is obtained by replacing the roles of $Q_{acc}$ and
$Q_{rej}$.
\end{proof}

Now, we further prove these results presented in Section 1.
Firstly we need a definition that is used in the following
results.

{\bf Definition 1.} We call 2qfa
$M=(Q,\Sigma,\delta,q_{0},Q_{acc},Q_{rej})$ {\it non-recurrent},
if for any $q\in Q$ with $q\not=q_{0}$,
$\delta(q,\sigma,q_{0},d)=0$ for any
$\sigma\in\Sigma\cup\{\cent,\$\}$.

Also, the well-formed conditions of 2qfa's given by Kodacs and
Watrous [17] for justifying the unitarity of evolution will be
used in what follows. Therefore we recall these conditions here. A
2qfa $M=(Q,\Sigma,\delta,q_{0},Q_{acc},Q_{rej})$ is well-formed of
and only if for any $\sigma,\sigma_{1},\sigma_{2}\in
\Sigma\cup\{\cent,\$\}$, and any $q_{1}q_{2}\in Q$, the following
hold.

1.
$\sum_{q^{'},d}\overline{\delta(q_{1},\sigma,q^{'},d)}\delta(q_{2},\sigma,q^{'},d)=\left\{
\begin{array}{ll} 1,&q_{1}=q_{2},\\
0,&q_{1}\not=q_{2},
\end{array}
\right.
$

2.
$\sum_{q^{'}}\left(\overline{\delta(q_{1},\sigma_{1},q^{'},1)}\delta(q_{2},\sigma_{2},q^{'},0)+
\overline{\delta(q_{1},\sigma_{1},q^{'},1)}\delta(q_{2},\sigma_{2},q^{'},-1)\right)=0$,

3.
$\sum_{q^{'}}\overline{\delta(q_{1},\sigma_{1},q^{'},1)}\delta(q_{2},\sigma_{2},q^{'},-1)=0$.

{\bf Theorem 1.} For any languages $L_{1}$ and $L_{2}$ over
$\Sigma_{1}$ and $\Sigma_{2}$ respectively, and any $n\in {\it
N}$, let $M_{1}$ and $M_{2}$ be the minimum 2qfa for
$L_{1}\cap\Sigma_{1}^{n}$ and $L_{2}\cap\Sigma_{2}^{n}$ with error
probabilities bounded by $\epsilon_{1}$ and $\epsilon_{2}$,
respectively. If $M_{2}$ is non-recurrent, then
\begin{equation}
QSC_{\epsilon}[L_{1}\cap L_{2}](n)\leq
QSC_{\epsilon_{1}}[L_{1}](n)+QSC_{\epsilon_{2}}[L_{2}](n)\times
(n+2)\times |Q_{acc,1}|-|Q_{acc,1}|
\end{equation}
\begin{equation}
QSC_{\epsilon}[L_{1}\cup L_{2}](n)\leq
QSC_{\epsilon_{1}}[L_{1}](n)+QSC_{\epsilon_{2}}[L_{2}](n)\times
(n+2)\times |Q_{rej,1}|-|Q_{rej,1}|
\end{equation}
where
$\epsilon=\epsilon_{1}+\epsilon_{2}-\epsilon_{1}\epsilon_{2}$,
$|Q_{acc,1}|$ denotes the number of the rejecting states of
$M_{1}$.

\begin{proof}
The basic idea for constructing 2qfa $M$ that accepts $L_{1}\cap
L_{2}$ with the length of input strings being $n$ is as follows:
Firstly let $M$ simulate $M_{1}$. If $M_{1}$ rejects the input
then the computation ends with rejection; otherwise, $M$ continues
to simulate $M_{2}$. Now we formally describe the process of
proof. Assume that 2qfa
$M_{i}=(Q_{i},\Sigma_{i},\delta_{i},q_{i0},Q_{acc,i},Q_{rej,i})$
accepts $L_{i}\cap\Sigma^{n}$ with error probability bounded at
most $\epsilon_{i}$, $i=1,2$, and $0\leq
\epsilon_{i}<\frac{1}{2}$. We construct
$M=(Q,\Sigma,\delta,q_{0},Q_{acc},Q_{rej})$ as follows.

$\Sigma=\Sigma_{1}\cap \Sigma_{2}$;

 $q_{0}=q_{10}$;

$Q=Q_{1}\cup Q_{2}^{'}\cup Q_{A}$ where

$Q_{2}^{'}=\bigcup_{i\in [0,n+1]_{{\bf Z}},q_{acc,1}\in
Q_{acc,1}}Q_{2}^{(i,q_{acc,1})}=[0,n+1]_{\bf Z}\times
Q_{acc,1}\times Q_{2}$, where

$Q_{2}^{(i,q_{acc,1})}=\{(i,q_{acc,1},q)|q\in Q_{2}\}$;

$Q_{acc}=\{(i,q_{acc,1},q_{acc})| q_{acc}\in Q_{acc,2},
i=0,1,\ldots,|Q_{acc,1}|,q_{acc,1}\in Q_{acc,1}\}$;

$Q_{rej}=Q_{rej,1}\cup \{(i,q_{acc,1},q_{rej})|q_{rej}\in
Q_{rej,2}, i=0,1,\ldots,|Q_{acc,1}|,q_{acc,1}\in Q_{acc,1}\}$;

$Q_{A}=\{(i,q_{acc,1}):q_{acc,1}\in Q_{acc,1},
i=0,1,2,\ldots,n+1\}$\\
where $Q_{A}$ is the set of auxiliary states that make the tape
head move back to the left end-marker $\cent$ when $M$ becomes a
state in $Q_{acc,1}$, and $(0,q_{acc,1})=q_{acc,1}$ for any
$q_{acc,1}\in Q_{acc,1}$. Furthermore, $\delta$ is defined as
follows: For any $\sigma\in\Sigma\cup \{\cent,\$\}$,
\newpage
\begin{eqnarray*}
&&\delta(q_{1},\sigma,q_{2},d)=\\
&&\\
 &&\left\{\begin{array}{ll} \delta_{1}(q_{1},\sigma,q_{2},d),&
q_{1},q_{2}\in
Q_{1},q_{1}\not\in Q_{acc,1},\\
0,& q_{1}\in Q_{acc,1},q_{2}\in Q_{1},\\
\delta_{2}(p_{1},\sigma,p_{2},d),&
q_{1}=(i,q_{acc,1},p_{1}),q_{2}=(i,q_{acc,1},p_{2})\in
Q_{2}^{(i,q_{acc,1})},i=0,1,2,\ldots,n+1,\\
1,&\sigma\not= \cent, q_{1}=(i,q_{acc,1}),q_{2}=(i+1,q_{acc,1}),d=-1,i=0,1,2,\ldots,n,\\
1,&\sigma= \cent,
q_{1}=(i,q_{acc,1}),q_{2}=(i,q_{acc,1},q_{20}),d=0,i=0,1,2,\ldots,n+1.\end{array}\right.
\end{eqnarray*}
Then, in terms of $\delta_{1}$ and $\delta_{2}$ it is ready to
extend $\delta$ such that it satisfies the well-formed conditions
of 2qfa's. Using $M$ to compute string $x\in
(\Sigma_{1}\cap\Sigma_{2})^{n}$, we obtain the following results:

(i) If $x\in L_{1}\cap L_{2}\cap \Sigma^{n}$, then $M$ accepts $x$
with probability at least
$(1-\epsilon_{1})(1-\epsilon_{2})=1-(\epsilon_{1}+\epsilon_{2}-\epsilon_{1}\epsilon_{2})$.

(ii) If $x\in \overline{L_{1}}\cap L_{2}\cap \Sigma^{n}$, or $x\in
\overline{L_{1}}\cap \overline{L_{2}}\cap \Sigma^{n}$, then $M$
rejects $x$ with probability at least $1-\epsilon_{1}$.

(iii) If $x\in L_{1}\cap \overline{L_{2}}\cap \Sigma^{n}$, then
$M$ rejects $x$ with probability
$(1-\epsilon_{1})(1-\epsilon_{2})$.

In addition, the number $|Q|$ of states of $M$ is
$QSC_{\epsilon_{1}}[L_{1}](n)+QSC_{\epsilon_{2}}[L_{2}](n)\times
(n+2)\times |Q_{acc,1}|-|Q_{acc,1}|$. Therefore, Eq. (6) is
proved.

The proof of Eq. (7) has certain similarity to Eq. (6). The 2qfa
$M$ for $(L_{1}\cup L_{2})\cap \Sigma^{n}$ can be constructed
according to the following process. For any $x\in \Sigma^{n}$,
firstly $M$ simulates $M_{1}$, and if $M_{1}$ accepts $x$ then $M$
also accepts $x$; otherwise $M$ continues to simulate $M_{2}$ and
the rest computation is then completed in terms of $M_{2}$.
Therefore, with the analogous idea as above, we construct
$M=(Q,\Sigma,\delta,q_{0},Q_{acc},Q_{rej})$ as follows.

$\Sigma=\Sigma_{1}\cup \Sigma_{2}$;

$q_{0}=q_{10}$;

$Q=Q_{1}\cup Q_{2}^{''}\cup Q_{B}$ where

$Q_{2}^{''}=\bigcup_{i\in [0,n+1]_{{\bf Z}},q_{rej,1}\in
Q_{rej,1}}Q_{2}^{(i,q_{rej,1})}=[0,n+1]_{\bf Z}\times
Q_{rej,1}\times Q_{2}$, where

$Q_{2}^{(i,q_{rej,1})}=\{(i,q_{rej,1},q)|q\in Q_{2}\}$;

$Q_{rej}=\{(i,q_{rej,1},q_{rej})| q_{rej}\in Q_{rej,2},
i=0,1,\ldots,|Q_{rej,1}|,q_{rej,1}\in Q_{rej,1}\}$;

$Q_{acc}=Q_{acc,1}\cup \{(i,q_{rej,1},q_{acc})|q_{acc}\in
Q_{acc,2}, i=0,1,\ldots,|Q_{rej,1}|,q_{rej,1}\in Q_{rej,1}\}$;

$Q_{B}=\{(i,q_{rej,1}):q_{rej,1}\in Q_{rej,1},
i=0,1,2,\ldots,n+1\}$\\
where $Q_{B}$ is the set of auxiliary states that make the tape
head move back to the left end-marker $\cent$ when $M$ becomes a
state in $Q_{rej,1}$, and $(0,q_{rej,1})=q_{rej,1}$ for any
$q_{rej,1}\in Q_{rej,1}$. Furthermore, $\delta$ is defined as
follows: For any $\sigma\in\Sigma\cup \{\cent,\$\}$,
\begin{eqnarray*}
&&\delta(q_{1},\sigma,q_{2},d)=\\
&&\left\{\begin{array}{ll} \delta_{1}(q_{1},\sigma,q_{2},d),&
q_{1},q_{2}\in
Q_{1},q_{1}\not\in Q_{rej,1},\sigma\in \Sigma_{1},\\
0,& q_{1}\in Q_{rej,1},q_{2}\in Q_{1},\sigma\in\Sigma_{1},\\
\delta_{2}(p_{1},\sigma,p_{2},d),&
q_{1}=(i,q_{rej,1},p_{1}),q_{2}=(i,q_{rej,1},p_{2})\in
Q_{2}^{(i,q_{rej,1})},\\ & i=0,1,2,\ldots,n+1,\sigma\in\Sigma_{2},\\
1,&\sigma\not=\cent, q_{1}=(i,q_{rej,1}),q_{2}=(i+1,q_{rej,1}),d=-1,i=0,1,2,\ldots,n,\\
1,&\sigma=\cent,
q_{1}=(i,q_{rej,1}),q_{2}=(i,q_{rej,1},q_{20}),d=0,i=0,1,2,\ldots,n+1.\end{array}\right.
\end{eqnarray*}

Also, $\delta$ can be extended to satisfy the well-formed
conditions of 2qfa's. Using $M$ to compute string $x\in
(\Sigma_{1}\cup\Sigma_{2})^{n}$, we obtain the following results:

(i) If $x\in L_{1}\cap \Sigma^{n}$, then $M$ accepts $x$ with
probability at least $1-\epsilon_{1}$.

(ii) If $x\in \overline{L_{1}}\cap L_{2}\cap \Sigma^{n}$, then $M$
rejects $x$ with probability at least
$(1-\epsilon_{1})(1-\epsilon_{2})$.

(iii) If $x\in \overline{L_{1}}\cap \overline{L_{2}}\cap
\Sigma^{n}$, then $M$ rejects $x$ with probability at least
$(1-\epsilon_{1})(1-\epsilon_{2})$.

\end{proof}

From Theorem 1 it follows the following corollary.

{\bf Corollary 1.} For any languages $L_{1}$ and $L_{2}$ over
$\Sigma_{1}$ and $\Sigma_{2}$ respectively, and any $n\in {\it
N}$, let $M_{1}$ and $M_{2}$ be the minimum 2qfa for
$L_{1}\cap\Sigma_{1}^{n}$ and $L_{2}\cap\Sigma_{2}^{n}$ with error
probabilities bounded by $\epsilon_{1}$ and $\epsilon_{2}$,
respectively. If the tape head of $M_{1}$ stays always at the left
end-marker when it enters accepting states, and
 $M_{2}$ is non-recurrent, then
\begin{equation}
QSC_{\epsilon}[L_{1}\cap L_{2}](n)\leq
QSC_{\epsilon_{1}}[L_{1}](n)+QSC_{\epsilon_{2}}[L_{2}](n)\times
|Q_{acc,1}|-|Q_{acc,1}|;
\end{equation}
if the tape head of $M_{1}$ stays always at the left end-marker
when it enters rejecting states, and
 $M_{2}$ is non-recurrent, then
\begin{equation}
QSC_{\epsilon}[L_{1}\cup L_{2}](n)\leq
QSC_{\epsilon_{1}}[L_{1}](n)+QSC_{\epsilon_{2}}[L_{2}](n)\times
 |Q_{rej,1}|-|Q_{rej,1}|
\end{equation}
where
$\epsilon=\epsilon_{1}+\epsilon_{2}-\epsilon_{1}\epsilon_{2}$,
$|Q_{acc,1}|$ denotes the number of the rejecting states of
$M_{1}$.

\begin{proof} It is straightforward by the proof of Theorem 1.
\end{proof}

To show that the above bounds are not tight, we verify the
following propositions.

{\bf Proposition 1.} For alphabet $\Sigma=\{a,b_{1},b_{2}\}$, let
$L=\{a^{n}b_{1}^{n}a^mb_{2}^m: n,m\geq 1\}$. Then $L$ is accepted
by 2qfa with one-sided error in linear time.

\begin{proof}
The idea borrows Proposition 2 of [17] in which Kodacs and Watrous
proved that non-regular language $\{a^{m}b^{m}|m\geq 1\}$ can be
accepted by 2qfa with one-sided error in linear time. Here, for
any $N\in {\bf N}$, we construct machine $M_{N}$ in terms of the
following idea. First we let machine $M_{N}$ check whether the
input is of form $a^{+}b_{1}^{+}a^{+}b_{2}^{+}$. If not, then
$M_{N}$ rejects it at once; otherwise, let the tape head of
$M_{N}$ stays at the right end-marker $\$$, and then check whether
or not the length of $b_{2}$ and $a$ in the right side equals. If
not, then $M_{N}$ rejects it with probability at least
$1-\frac{1}{N}$; otherwise, with probability one $M_{N}$ continues
to check the equality of the length of $b_{1}$ and $a$ in the left
side. If not, then $M_{N}$ rejects it with probability at least
$1-\frac{1}{N}$; otherwise, $M_{N}$ accepts it with probability
one. Now we give the formal description of
$M_{N}=(Q,\Sigma,\delta,q_{0},Q_{acc},Q_{rej})$ where the state
set $Q$ consists of the all states appearing in the following,

$\delta(q,\sigma,p,d)=\left\{\begin{array}{ll} \langle
p|V_{\sigma}|q\rangle,& D(p)=d,\\
0,& D(p)\not=d,\end{array}\right.$ and $Q_{acc}=\{s_{N}^{(2)}\}$;

$Q_{rej}=\{q_{r}^{(0)},q_{r}^{(1)},q_{r}^{(2)},q_{r}^{(3)}\}\cup\bigcup_{i=1}^{2}\{s_{1}^{(i)},s_{2}^{(i)},\ldots,s_{N-1}^{(i)}\}$;

For all $\sigma\in\Sigma\cup\{\cent,\$\}$, unitary operators
$V_{\sigma}$ on $\ell_{2}(Q)$ are defined as follows:

\[
\begin{array}{ll}
V_{\cent}|q_{0}\rangle=|q_{0}\rangle,&V_{\cent}|q_{1}\rangle=|q_{r}^{(1)}\rangle,\\
V_{\cent}|r_{j,0}^{(2)}\rangle=\frac{1}{\sqrt{N}}\sum_{l=1}^{N}
\exp (\frac{2\pi i}{N}jl)|s_{l}^{(2)}\rangle, 1\leq j\leq N,&\\
&\\
V_{\$}|q_{0}\rangle=|q_{r}^{(0)}\rangle,&V_{\$}|q_{1}\rangle=|q_{r}^{(1)}\rangle,\\
V_{\$}|q_{2}\rangle=|q_{r}^{(2)}\rangle,&V_{\$}|q_{4}\rangle=|q_{r}^{(3)}\rangle,\\
V_{\$}|q_{6}\rangle=\frac{1}{\sqrt{N}}\sum_{j=1}^{N}|r_{j,0}^{(1)}\rangle,&\\
&\\
V_{a}|q_{0}\rangle=|q_{0}\rangle,&
V_{a}|q_{1}\rangle=|q_{2}\rangle,\\
V_{a}|q_{2}\rangle=|q_{3}\rangle,& V_{a}|q_{4}\rangle=|q_{4}\rangle,\\
V_{a}|q_{5}\rangle=|q_{6}\rangle,&
V_{a}|q_{6}\rangle=|q_{r}^{(0)}\rangle,\\
V_{a}|s_{N}^{(1)}\rangle=\frac{1}{\sqrt{N}}\sum_{j=1}^{N}|r_{j,0}^{(2)}\rangle,&
V_{a}|r_{j,0}^{(2)}\rangle=|r_{j,j}^{(2)}\rangle, 1\leq j\leq N,\\
V_{a}|r_{j,k}^{(2)}\rangle=|r_{j,k-1}^{(2)}\rangle, 1\leq j\leq N,
1\leq k\leq j, & V_{a}|r_{j,0}^{(1)}\rangle=|r_{j,j}^{(1)}\rangle,
1\leq j\leq N,\\
V_{a}|r_{j,k}^{(1)}\rangle=|r_{j,k-1}^{(1)}\rangle, 1\leq j\leq
N,1\leq k\leq j,&\\
&\\
V_{b_{1}}|q_{0}\rangle=|q_{1}\rangle,&V_{b_{1}}|q_{2}\rangle=|q_{2}\rangle,\\
V_{b_{1}}|q_{3}\rangle=|q_{4}\rangle,&V_{b_{1}}|q_{4}\rangle=|q_{r}^{(0)}\rangle,\\
V_{b_{1}}|q_{6}\rangle=|q_{r}^{(1)}\rangle, 1\leq j\leq
N,&V_{b_{1}}|r_{j,0}^{(2)}\rangle=|r_{j,N-j+1}^{(2)}\rangle, 1\leq
j\leq N,\\
V_{b_{1}}|r_{j,0}^{(1)}\rangle=\frac{1}{\sqrt{N}}\sum_{l=1}^{N}
\exp (\frac{2\pi i}{N}jl)|s_{l}^{(1)}\rangle, 1\leq j\leq
N,&\\
V_{b_{1}}|r_{j,k}^{(2)}\rangle=|r_{j,k-1}^{(2)}\rangle, 1\leq
j\leq N,1\leq k\leq N-j+1,&\\
&

\end{array}
\]

\[
\begin{array}{ll}

V_{b_{2}}|q_{0}\rangle=|q_{r}^{(0)}\rangle,&
V_{b_{2}}|q_{4}\rangle=|q_{5}\rangle,\\
V_{b_{2}}|q_{6}\rangle=|q_{6}\rangle,&
V_{b_{2}}|r_{j,0}^{(1)}\rangle=|r_{j,N-j+1}^{(1)}\rangle, 1\leq
j\leq N,\\
V_{b_{2}}|r_{j,k}^{(1)}\rangle=|r_{j,k-1}^{(1)}\rangle, 1\leq
j\leq N,1\leq k\leq N-j+1,&\\
&\\
D(q_{0})=+1,&D(r_{j,0}^{(i)})=-1,\\
D(q_{1})=-1,&D(r_{j,k}^{(i)})=0,i=1,2, 1\leq j\leq N, k\not=0,\\
D(q_{2})=+1,&D(q_{3})=-1,\\
D(q_{4})=+1,&D(q_{5})=-1,\\
D(q_{6})=+1,& D(s_{l}^{(1)})=0,1\leq l\leq N-1,\\
D(s_{2}^{(1)})=0,1\leq l\leq N,&D(s_{N}^{(1)})=+1,\\
D(q_{r}^{(i)})=0,i=0,1,2,3.&

\end{array}
\]

\end{proof}

{\bf Proposition 2.} For alphabet $\Sigma=\{a,b_{1},b_{2}\}$, let
$L_{1}=\{a^{+}b_{1}^{+}a^mb_{2}^m: m\geq 1\}$ and
$L_{2}=\{a^{n}b_{1}^{n}a^+b_{2}^+: n\geq 1\}$. Then there exist
2qfa $M_{1}$ and $M_{2}$ accepting $L_{1}$ and $L_{2}$,
respectively, with one-sided error in linear time.

\begin{proof}
By changing the construction of $M_{N}$ as the proof of
Proposition 1, we can obtain 2qfa's $M_{1}$ and $M_{2}$ to accept
$L_{1}$ and $L_{2}$ with one-sided error, respectively. In the
interest of completeness, we give the detailed construction of
$M_{1}=(Q_{1},\Sigma,\delta_{1},q_{0},Q_{acc},Q_{rej})$ and
$M_{2}=(Q_{2},\Sigma,\delta_{2},q_{0},Q_{acc},Q_{rej})$. For
$M_{1}$,

$\delta_{1}(q,\sigma,p,d)=\left\{\begin{array}{ll} \langle
p|V_{\sigma}|q\rangle,& D(p)=d,\\
0,& D(p)\not=d,\end{array}\right.$ and $Q_{acc}=\{s_{N}\}$;

$Q_{rej}=\{q_{r}^{(0)},q_{r}^{(1)},q_{r}^{(2)},q_{r}^{(3)}\}\cup\{s_{1},s_{2},\ldots,s_{N-1}\}$;

For all $\sigma\in\Sigma\cup\{\cent,\$\}$, unitary operators
$V_{\sigma}$ on $\ell_{2}(Q)$ are defined as follows:

\[
\begin{array}{ll}
V_{\cent}|q_{0}\rangle=|q_{0}\rangle,&V_{\cent}|q_{1}\rangle=|q_{r}^{(1)}\rangle,\\
&\\
V_{\$}|q_{0}\rangle=|q_{r}^{(0)}\rangle,&V_{\$}|q_{1}\rangle=|q_{r}^{(1)}\rangle,\\
V_{\$}|q_{2}\rangle=|q_{r}^{(2)}\rangle,&V_{\$}|q_{4}\rangle=|q_{r}^{(3)}\rangle,\\
V_{\$}|q_{6}\rangle=\frac{1}{\sqrt{N}}\sum_{j=1}^{N}|r_{j,0}\rangle,&\\

\end{array}
\]
\[
\begin{array}{ll}

V_{a}|q_{0}\rangle=|q_{0}\rangle,&
V_{a}|q_{1}\rangle=|q_{2}\rangle,\\
V_{a}|q_{2}\rangle=|q_{3}\rangle,& V_{a}|q_{4}\rangle=|q_{4}\rangle,\\
V_{a}|q_{5}\rangle=|q_{6}\rangle,&
V_{a}|q_{6}\rangle=|q_{r}^{(0)}\rangle,\\
V_{a}|r_{j,0}\rangle=|r_{j,j}\rangle, 1\leq j\leq N,&
V_{a}|r_{j,k}\rangle=|r_{j,k-1}\rangle, 1\leq j\leq N, 1\leq k\leq
j, \\
V_{a}|r_{j,0}\rangle=|r_{j,j}\rangle,
1\leq j\leq N,&\\
&\\
V_{b_{1}}|q_{0}\rangle=|q_{1}\rangle,&V_{b_{1}}|q_{2}\rangle=|q_{2}\rangle,\\
V_{b_{1}}|q_{3}\rangle=|q_{4}\rangle,&V_{b_{1}}|q_{4}\rangle=|q_{r}^{(0)}\rangle,\\
V_{b_{1}}|q_{6}\rangle=|q_{r}^{(1)}\rangle, 1\leq j\leq
N,&\\
V_{b_{1}}|r_{j,0}\rangle=\frac{1}{\sqrt{N}}\sum_{l=1}^{N} \exp
(\frac{2\pi i}{N}jl)|s_{l}\rangle, 1\leq j\leq N,&\\
&\\

V_{b_{2}}|q_{0}\rangle=|q_{r}^{(0)}\rangle,&
V_{b_{2}}|q_{4}\rangle=|q_{5}\rangle,\\
V_{b_{2}}|q_{6}\rangle=|q_{6}\rangle,&
V_{b_{2}}|r_{j,0}\rangle=|r_{j,N-j+1}\rangle, 1\leq
j\leq N,\\
V_{b_{2}}|r_{j,k}\rangle=|r_{j,k-1}\rangle, 1\leq
j\leq N,1\leq k\leq N-j+1,&\\
&\\
D(q_{0})=+1,&D(r_{j,0})=-1,\\
D(q_{1})=-1,&D(r_{j,k})=0, 1\leq j\leq N, k\not=0,\\
D(q_{2})=+1,&D(q_{3})=-1,\\
D(q_{4})=+1,&D(q_{5})=-1,\\
D(q_{6})=+1,& D(s_{l})=0,1\leq l\leq N-1,\\
D(q_{r}^{(i)})=0,i=0,1,2,3,&D(s_{N})=+1.

\end{array}
\]
Therefore,
$Q_{1}=\{q_{0},q_{1},\ldots,q_{6}\}\cup\{s_{1},\ldots,s_{N}\}\cup\{q_{r}^{(0)},q_{r}^{(1)},q_{r}^{(2)},q_{r}^{(3)}\}$,
in which $Q_{acc}=\{s_{N}\}$,
$Q_{rej}=\{s_{1},\ldots,s_{N-1}\}\cup\{q_{r}^{(0)},q_{r}^{(1)},q_{r}^{(2)},q_{r}^{(3)}\}$.

Next we construct
$M_{2}=(Q_{2},\Sigma,\delta_{2},q_{0},Q_{acc},Q_{rej})$ which is
largely similar to $M_{1}$. We still present the detailed
definitions of $V_{\sigma}$ for any
$\sigma\in\Sigma\cup\{\cent,\$\}$.

\[
\begin{array}{ll}
V_{\cent}|q_{0}\rangle=|q_{0}\rangle,&V_{\cent}|q_{1}\rangle=|q_{r}^{(1)}\rangle,\\
V_{\cent}|r_{j,0}\rangle=\frac{1}{\sqrt{N}}\sum_{l=1}^{N} \exp
(\frac{2\pi i}{N}jl)|s_{l}\rangle, 1\leq j\leq N,&\\
&\\
V_{\$}|q_{0}\rangle=|q_{r}^{(0)}\rangle,&V_{\$}|q_{1}\rangle=|q_{r}^{(1)}\rangle,\\
V_{\$}|q_{2}\rangle=|q_{r}^{(2)}\rangle,&V_{\$}|q_{4}\rangle=|q_{r}^{(3)}\rangle,\\
V_{\$}|q_{6}\rangle=|q_{7}\rangle,&\\

\end{array}
\]

\[
\begin{array}{ll}

V_{a}|q_{0}\rangle=|q_{0}\rangle,&
V_{a}|q_{1}\rangle=|q_{2}\rangle,\\
V_{a}|q_{2}\rangle=|q_{3}\rangle,& V_{a}|q_{4}\rangle=|q_{4}\rangle,\\
V_{a}|q_{5}\rangle=|q_{6}\rangle,&
V_{a}|q_{6}\rangle=|q_{r}^{(0)}\rangle,\\
V_{a}|r_{j,0}\rangle=|r_{j,j}\rangle, 1\leq j\leq N,&
V_{a}|r_{j,k}\rangle=|r_{j,k-1}\rangle, 1\leq j\leq N, 1\leq k\leq
j, \\
V_{a}|r_{j,0}\rangle=|r_{j,j}\rangle,
1\leq j\leq N,&V_{a}|q_{7}\rangle=\frac{1}{\sqrt{N}}\sum_{j=1}^{N}|r_{j,0}\rangle, \\
&\\
V_{b_{1}}|q_{0}\rangle=|q_{1}\rangle,&V_{b_{1}}|q_{2}\rangle=|q_{2}\rangle,\\
V_{b_{1}}|q_{3}\rangle=|q_{4}\rangle,&V_{b_{1}}|q_{4}\rangle=|q_{r}^{(0)}\rangle,\\
V_{b_{1}}|q_{6}\rangle=|q_{r}^{(1)}\rangle, 1\leq j\leq
N,&V_{b_{1}}|r_{j,0}\rangle=|r_{j,N-j+1}\rangle, 1\leq
j\leq N,\\
V_{b_{1}}|r_{j,k}\rangle=|r_{j,k-1}\rangle, 1\leq j\leq N,1\leq
k\leq j,&\\
&\\

V_{b_{2}}|q_{0}\rangle=|q_{r}^{(0)}\rangle,&
V_{b_{2}}|q_{4}\rangle=|q_{5}\rangle,\\
V_{b_{2}}|q_{6}\rangle=|q_{6}\rangle,&
V_{b_{2}}|r_{j,0}\rangle=|r_{j,N-j+1}\rangle, 1\leq
j\leq N,\\
V_{b_{2}}|r_{j,k}\rangle=|r_{j,k-1}\rangle, 1\leq
j\leq N,1\leq k\leq N-j+1,&\\
&\\
D(q_{0})=+1,&D(r_{j,0})=-1,\\
D(q_{1})=-1,&D(r_{j,k})=0, 1\leq j\leq N, k\not=0,\\
D(q_{2})=+1,&D(q_{3})=-1,\\
D(q_{4})=+1,&D(q_{5})=-1,\\
D(q_{6})=+1,& D(s_{l})=0,1\leq l\leq N-1,\\
D(q_{r}^{(i)})=0,i=0,1,2,3,&D(s_{N})=+1,\\
D(q_{7})=-1.&
\end{array}
\]

{\bf Remark 1.} From the above Propositions it follows that
$QSC_{1/N}(L_{1})\leq \frac{N(N+5)}{2}+10$, $QSC_{1/N}(L_{2})\leq
\frac{N(N+5)}{2}+11$, and $QSC_{1/N}(L_{1}\cap L_{2})\leq
\frac{N(2N+8)}{2}+10$, where $L_{1}\cap L_{2}=L$. This result
shows that the bound in Eq. (6) is not tight. Since $L_{1}\cup
L_{2}=\{a^{+}b_{1}^{+}a^{+}b_{2}^{+}\}$ is a regular language, the
bound in Eq. (7) is not tight either.

\end{proof}

Next we deal with the reversal of languages accepted by 2qfa's, by
demonstrating Theorem 2.

{\bf Theorem 2.} For any language $L$ over $\Sigma$, let $M$ be
the minimum 2qfa for $L_{1}$ with error probability bounded by
$\epsilon$. If $M$ is non-recurrent, then

\begin{proof}
With the condition that $M$ is non-recurrent, we only
need add a state $q_{0}^{'}$ as starting state for constructing a
2qfa $M^{R}$ for $L^{R}$, and let $q_{0}^{'}$ change to $q_{0}$
with its tape head moving to the right end-marker $\$$. Then
$M^{R}$ simulate $M$ in the reversal direction. Formally
$M^{R}=(Q\cup\{q_{0}^{'}\},\Sigma,\delta^{R},q_{0}^{'},Q_{acc},Q_{rej})$
where $q_{0}^{'}\not\in Q$, $\delta^{R}$ is defined as follows:

1) $\delta^{R}(q_{0}^{'},\cent, q_{0},-1)=1$;

2) $\delta^{R}(q,\$,p,d)=\delta(q,\cent,p,-d)$ for any $q,p\in Q$;

3) $\delta^{R}(q,\cent,p,d)=\delta(q,\$,p,-d)$ for any $q,p\in Q$;

4) $\delta^{R}(q,\sigma,p,d)=\delta(q,\sigma,p,-d)$ for any
$q,p\in Q$ and any $\sigma\in\Sigma$.

Then it is clear that $\delta^{R}$ satisfies the well-formed
conditions of 2qfa's if $\delta$ does. Therefore, the second
inequality has been proved. The first inequality is only an
inference from the second one, since $L=(L^{R})^{R}$.

\end{proof}

Finally we deal with catenation operation of 2qfs's. For technical
reason, we restrict the 2qfa's to be non-circular, that is, when a
machine's tape head is scanning the left end-marker (or the right
end-marker), the machine will not move its tape head left (right).
Also, in the interest of simplicity, we consider only the
languages without $\epsilon$.

In the interest of simplicity, as in Corollary 1, we would like to
assume that the tape head of the first machine $M_{1}$ stays at
the right end-marker when it enters accepting states. Without this
assumption one can also cope with it by virtue of the similar way
used in the proof of Theorem 1.

{\bf Theorem 3.} Let $L_{i}$ be a language over alphabet
$\Sigma_{i}$ with $\epsilon\not\in L_{i}$ for $i=1,2$. If
$\Sigma_{1}\cap\Sigma_{2}=\emptyset$, and the error probabilities
of the minimum 2qfs's $M_{1}$ and $M_{2}$ accepting $L_{1}$ and
$L_{2}$ are respectively $\epsilon_{1}$ and $\epsilon_{2}$, then
$L_{1}L_{2}$ is accepted by a 2qfa $M$ with error probability
$\epsilon=\epsilon_{1}+\epsilon_{2}-\epsilon_{1}\epsilon_{2}$.

\begin{proof}
Firstly we let $M$ check whether or not the input is of the form
$\Sigma^{+}\Sigma^{+}$. If not, $M$ rejects it immediately;
otherwise, $M$ simulate $M_{1}$. If the input is rejected, then
$M$ rejects it also; otherwise $M$ continues to simulate $M_{2}$
for the second part of the input, and therefore $M_{2}$ determines
the accepting or rejecting result. Specifically,
$M=(Q,\Sigma_{1}\cup\Sigma_{2},\delta,q_{0},Q_{acc},Q_{rej})$
where $\delta$ is defined as follows.
\[
\begin{array}{ll}
\delta(q_{0},\cent,q_{0},1)=1,& \delta(q_{0},\sigma,q_{0},1)=1,
\sigma\in
\Sigma_{1},\\
\delta(q_{0},\sigma,q_{1},-1)=1,
\sigma\in\Sigma_{2},&\delta(q_{2},\$,q_{3},-1)=1,\\
\delta(q_{1},\cent,q_{r},0)=1,&\delta(q_{3},\sigma,q_{3},-1)=1,\sigma\in\Sigma_{2},\\
\delta(q_{1},\sigma,q_{2},1)=1,
\sigma\in\Sigma_{1},&\delta((q_{3},\sigma,q_{3},-1)=1,\sigma\in\Sigma_{1},\\
\delta(q_{2},\sigma,q_{2},1)=1,\sigma\in\Sigma_{2},&\delta(q_{3},\cent,q_{1,0},0)=1,\\
\delta(q_{2},\sigma,q_{r},0)=1,\sigma\in\Sigma_{1}.&

\end{array}
\]
The above process checks whether the input is the form of
$\Sigma_{1}^{+}\Sigma_{2}^{+}$. If it is, $M$ begins with
simulating $M_{1}$. Therefore, $\delta$ is further defined as
follows.
\[
\delta(p_{1},\sigma,p_{2},d)=\\
\left\{\begin{array}{ll} \delta_{1}(p_{1},\sigma,p_{2},d),&
\sigma\not=\cent,\sigma\in\Sigma_{1},p_{1},p_{2}\in Q_{1},\\
\delta_{1}(p_{1},\$,p_{2},d),& \sigma\in\Sigma_{2},p_{1},p_{2}\in
Q_{1},\\
1,& \sigma\in\Sigma_{2},p_{1}\in Q_{acc,1},p_{2}=p_{2,0},d=-1,\\
\delta_{2}(p_{1},\cent,p_{2},d),&\sigma\in\Sigma_{1},p_{1}=p_{2,0},p_{2}\in
Q_{2},\\
\delta_{2}(p_{1},\sigma,p_{2},d),&\sigma\in\Sigma_{2}\cup\{\$\},p_{1},p_{2}\in
Q_{2}.
\end{array}
\right.
\]

It is seen that $Q=Q_{1}\cup
Q_{2}\cup\{q_{0},q_{1},q_{2},q_{3},q_{r}\}$, where
$Q_{acc}=Q_{acc,1}$, and $Q_{rej}=\{q_{r}\}\cup Q_{rej,1}\cup
Q_{rej,2}$. Then, $M$ accepts $x\in L_{1}L_{2}$ with probability
at least $(1-\epsilon_{1})(1-\epsilon_{2})$, and rejects $x\not\in
L_{1}L_{2}$ with probability at least
$(1-\epsilon_{1})(1-\epsilon_{2})$.
\end{proof}

Finally we present an example to show that the conditions such as
$\Sigma_{1}\cap\Sigma_{2}=\emptyset$ in Theorem 3 are not
necessary.

{\bf Example.} For $\Sigma_{1}=\{a,b_1\}$ and
$\Sigma_{2}=\{a,b_{2}\}$, languages $L_{1}$ over $\Sigma_{1}$ and
$L_{2}$ over $\Sigma_{2}$ are respectively defined as:

$L_{1}=\{a^{m}b_{1}^{m}|m\geq 1\}$ and

$L_{2}=\{a^{m}b_{2}^{m}|m\geq 1\}$.

In terms of Proposition 2 of [17], $L_{1}$ and $L_{2}$ can be
accepted by 2qfa's with one-sided error in linear time. As well,
by Proposition 1 above, the catenation of $L_{1}$ and $L_{2}$ as
the language over $\Sigma_{1}\cup\Sigma_{2}$
\[L_{1}L_{2}=\{a^{m}b_{1}^{m}a^{m}b_{2}^{m}|m\geq 1\}\]
is accepted by 2qfa  with one-sided error in linear time.
Therefore, this shows that the conditions such as
$\Sigma_{1}\cap\Sigma_{2}=\emptyset$ in Theorem 3 are not
necessary.  $\Box$

\section*{5. Concluding remarks and future works}

In this report, we dealt with the state complexity of some
operations (including complementation, intersection, union,
reversals, catenation) on two-way quantum finite automata. We
proved a number of upper bounds of the size of states for these
operations, and also obtained lower bound for reversal operation.
Also, we provided in detail a number of non-regular languages and
demonstrated that these languages can be accepted by two-way
quantum finite automata with one-sided error probabilities in
linear time. In terms of these examples we have seen that the
bounds obtained for these operations are not tight. Therefore,
this motivates to further consider related issues along this
direction.

Therefore, the further work is how to improve these bounds to make
them optimum, and how to verify related lower bounds for
intersection and union. In particular, there are some restricted
conditions in these theorems (such as non-recurrent), so, proving
these theorems without these conditions is worth further
exploring.

As is well-known, classical interactive proof systems [8,11] have
played an important role in the study of computational complexity,
and have been successfully applied to cryptography systems.
Notably, by generalizing the classical interactive proof systems
of Dwork and Stockmeyer [11] to quantum framework,  Nishimura and
Yamakami [22] recently have significantly dealt with quantum
interactive proof systems by using 2qfa's. Furthermore, to study
zero-knowledge quantum interactive proof systems by using two-way
quantum finite automata as verifiers is a significant issue for
the future consideration.

\section*{Acknowledgment}

I would like to thank Dr. Tomoyuki Yamakami for helpful discussion
regarding qfa's and this report was motivated by the discussion.

\end{document}